\documentclass[aps,prb,postprint,groupedaddress,twocolumn,showpacs,10pt]{revtex4-1}
\usepackage{graphicx}
\usepackage{epsfig}
\usepackage{epstopdf}
\usepackage{subfigure}
\usepackage{appendix}
\usepackage{relsize}
\DeclareMathAlphabet\mathbfcal{OMS}{cmsy}{b}{n}
\usepackage{amsmath,amssymb}
\usepackage{ulem}
\usepackage{color}
\begin{document}
\title{Bilayer graphene in strong ultrafast laser fields}
\author{Pardeep Kumar}
%\email[]{}
\author{Thakshila M. Herath}
\author{Vadym Apalkov}
\author{Mark I. Stockman}
\affiliation{Center for Nano-Optics (CeNO) and Department of Physics and Astronomy,
Georgia State University, Atlanta, Georgia 30303, USA}

\date{\today}

\begin{abstract}
We theoretically investigate the interaction of an ultrastrong femtosecond-long linearly polarized optical pulse with AB-stacked bilayer graphene. The pulse excite electrons from the valence into the conduction band, resulting in finite conduction band population. Such a redistribution of electrons results in the generation of current which can be manipulated by the angle of incidence of the pulse. For the normal incidence, the current along a direction transverse to the polarization of the optical pulse is zero. However, the interlayer symmetry is broken up by a finite angle of incidence which causes induction of electric current in the direction perpendicular to the $x$-$z$ plane  of polarization of the pulse.  We show that the magnitude and the direction of such a current as well as charge transfer along this direction can be manipulated by tuning the angle of incidence of the laser pulse. Further, the symmetry of the system prohibits the generation of transverse current if the pulse is polarized along $y$-direction.  
\end{abstract}


\maketitle

\section{Introduction}
Graphene, a two dimensional (2D) material with honeycomb lattice structure, is a promising candidate for future technology \cite{neto2009}. The uniqueness of graphene is due to its remarkable physical properties \cite{abergel2010}: the valence band (VB) and the conduction band (CB) are touching each other at the Dirac points thereby making it a gapless semiconductor \cite{geim2007}. It is chiral and is characterized by the  Berry phase of $\pm\pi$ at the Dirac points \cite{liu2011,xiao2010,mikitik1999}. Interestingly, the electron dynamics in graphene is governed by a massless Dirac equation \cite{semenoff1984}. This unusual electronic behavior gives rise to various unprecedented phenomenona, such as half-integer quantum Hall effect \cite{zhang2005,novoselov2005,kane2005}, ballistic transport \cite{baringhaus2014}, and Klein tunneling due to the absence of back scattering \cite{young2009}.  

Interestingly, multilayer graphene systems come to demand for their increased electrical or thermal properties \cite{castro2007,gong2012} and optical signatures \cite{yan2012}.  The simplest and thinnest intercalated  structure of interest is the bilayer graphene \cite{mccann2013,rozhkova2016}. Despite of having many properties similar to monolayer graphene \cite{neto2009}, the bilayer graphene  is potentially different in terms of underlying features: (i) in bilayer graphene the dispersion near the Dirac points is parabolic \cite{mccann2006} unlike the monolayer graphene where it is linear; (ii) the charge carriers in bilayer graphene are \textit{massive} chiral quasiparticles \cite{li2007} rather than massless ones in monolayer graphene; (iii) the Berry phase in bilayer \cite{novoselov2006} is 2$\pi$  while in monolayer it is $\pi$; (iv)  each individual layer in bilayer graphene can be tuned separately by doping \cite{ohta2006} or gating \cite{mccann2006_v2,oostinga2007}. Moreover, the perpendicular electric field  removes the interlayer symmetry \cite{zhang2010} by introducing the difference in the on-site energies, which opens a band gap at the $K$ and $K^{\prime}$  points. The band gap can be tuned by the magnitude of transverse electric field. This distinctive feature of bilayer grahene makes it a potential candidate for electronic applications, which in monolayer graphene are limited due to its semimetal nature. 

Recent advances in ultrafast laser technology have provided a versatile platform to explore the coherent control of electron dynamics at the sub-femtosecond time scale   \cite{krausz2014}. In particular, the interaction of strong ultrafast laser pulses with 2D materials opens a pathway to probe their extremely nonlinear behavior \cite{ghimire2014,nematollahi2018}. Such optical pulses produce dramatic changes in the electron dynamics of graphene \cite{heide2018}. For instance, linearly polarized pulses produce interference fringes in the CB population distribution in the reciprocal space of graphene. This effect is due to quantum interference caused by a double passage by an electron of the Dirac points \cite{kelardeh2015} during the pulse. Consequently, in  graphene there exist a current along the direction of the polarization of the pulse . However, the interaction of linearly polarized pulse with gapped Dirac materials  leads to the generation of an ultrafast  current in the direction transverse to the plane of polarization of the pulse \cite{azar20192}. Such ultrafast  electric currents have been observed experimentally \cite{higuchi2017}.    Contrary to linearly polarized pulses, the chiral optical pulses selectively populate the $K$ and $K^{\prime}$ valleys \cite{kelardeh2016}, thereby generating valley polarization in gapped Dirac materials \cite{azar20182,azar2019}. Moreover, these effects are attributed to the existence of topological resonance which arises due to competition between dynamic and topological phase \cite{azar20182}. 
%It was also theoretically predicted that for gapped graphene, the ultrafast circularly polarized pulse of just a single oscillation significantly populates one valley with respect to the other one, depending on the chirality of the pulse \cite{azar2019}.  This gives rise to a large valley polarization, an effect similar to one predicted in transition metal dichalcogenides (TMDC) \cite{azar20182}. 

In this paper, we study the interaction of AB-stacked bilayer graphene with an ultrafast optical pulse. The laser pulse is applied at an oblique incidence and the normal component of it results in the interlayer asymmetry which opens a dynamical band gap at the $K$ and $K^{\prime}$ points. Unlike monolayer, the bilayer graphene in AB-stacking is axially symmetric only about $y$-axis and not along $x$-axis. We show that if the pulse is polarized along $x$-direction then it leads to the generation of a transverse electric current which can be maneuvered by means of the angle of incidence. Such a transverse current also results in the finite charge transfer. However, symmetry consideration of the system forbids any such transverse current for $y$-polarization of the pulse. We also show that the ultrafast electron dynamics in bilayer graphene remains nonadiabatic and irreversible, likewise in graphene, even when the interlayer symmetry is broken by the normal component of the applied pulse.

The paper is organized as follows. In Sec. \ref{model}, we describe the model of bilayer graphene and introduce the main equations. In Sec. \ref{results}, we present and discuss our main results. Finally, the concluding remarks are given in Sec. \ref{conclusion}.

\section{Model}
\label{model}
\begin{figure}[ht!]
\begin{center}
\begin{tabular}{cc}
\includegraphics[scale=0.35]{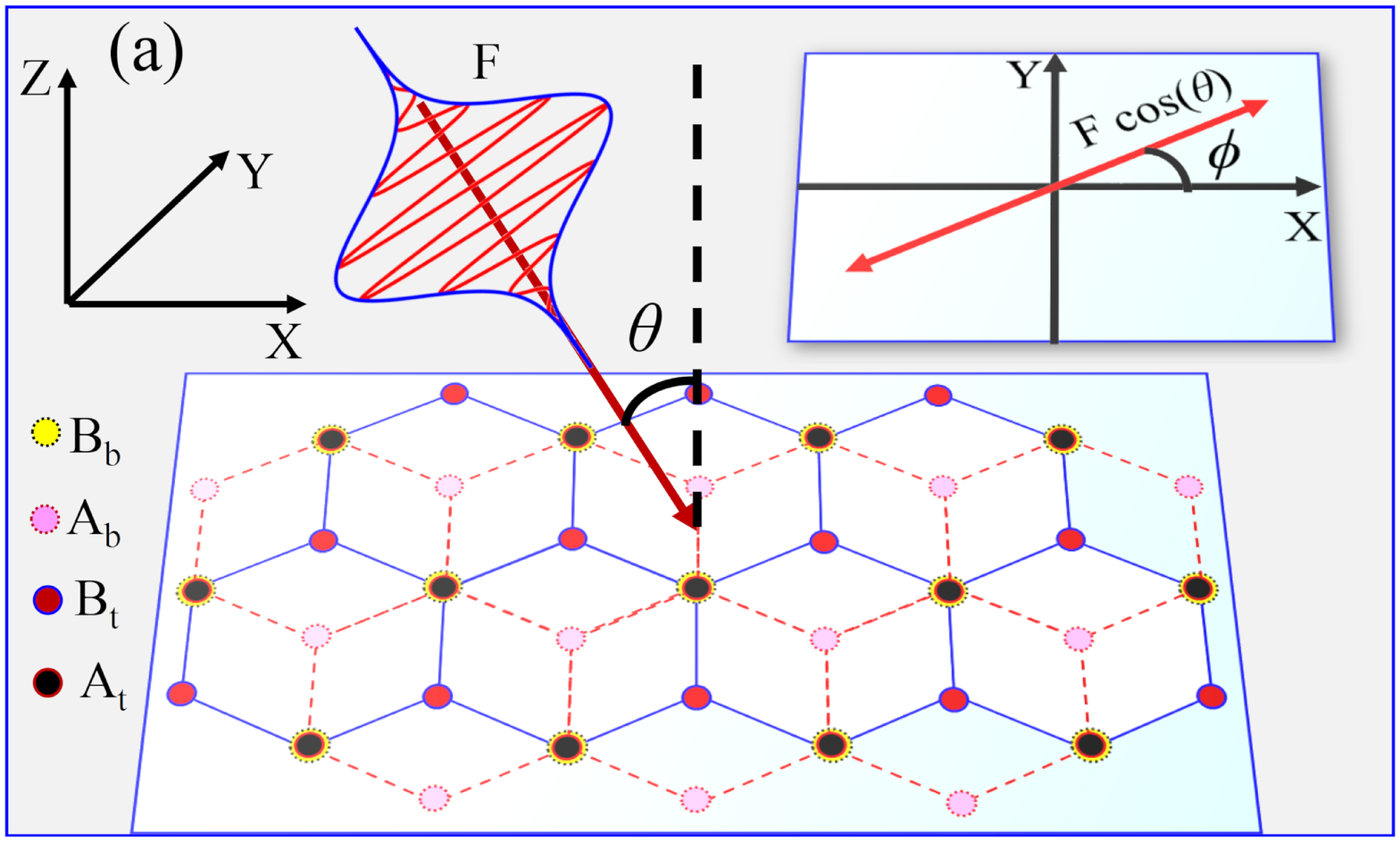}\\
\includegraphics[scale=0.3]{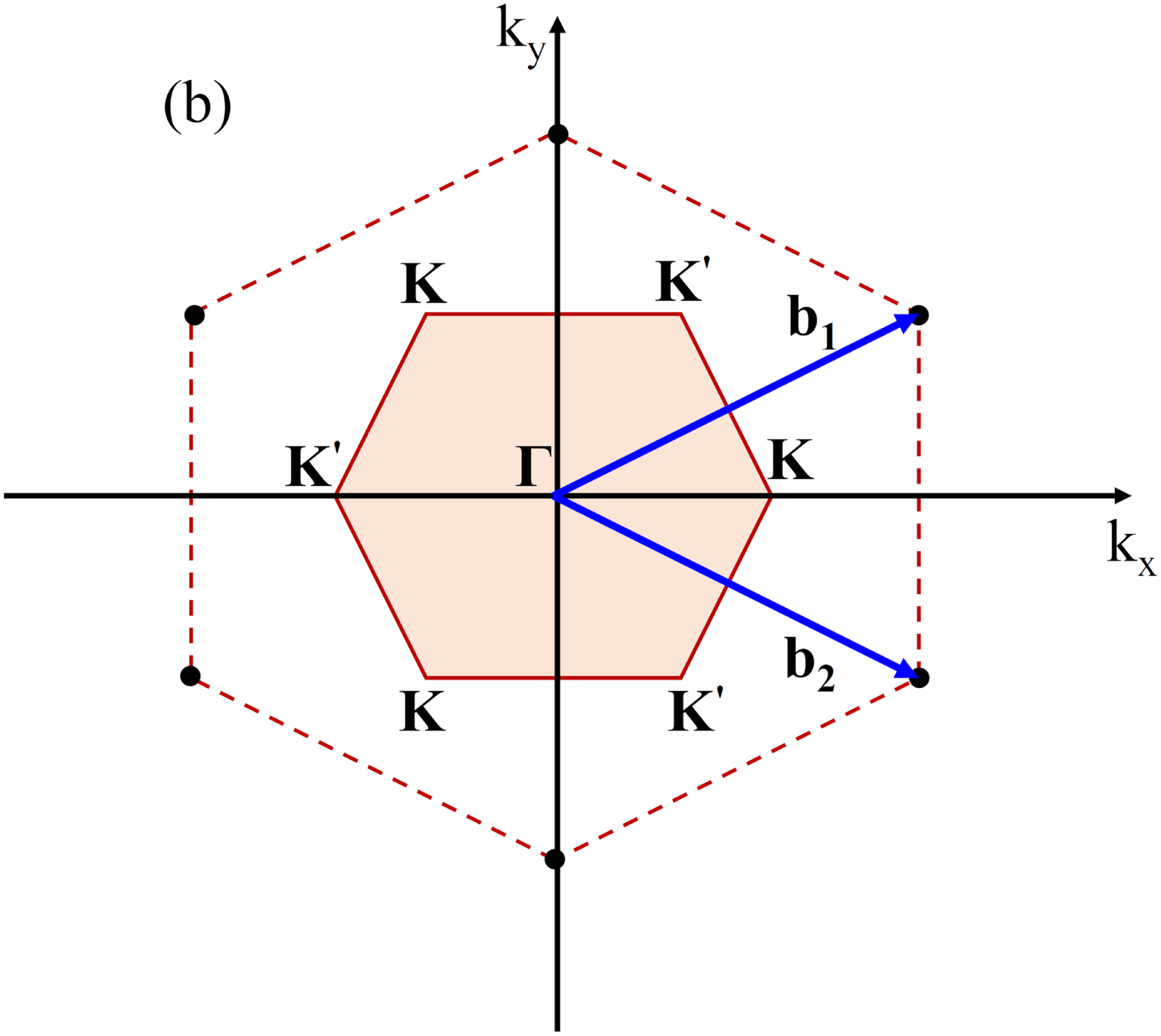}
\end{tabular}
\caption{(a) AB-stacking of bilayer graphene consisting of two coupled monolayers of graphene with honeycomb crystal structure. The bottom layer is represented by dashed red lines and the top layer - by solid blue lines. Each carbon atom in  sublattice $B_{b}$ of the bottom layer is exactly below the carbon atoms in sublattice $A_{t}$ of the top layer. The bilayer graphene interacts with the ultrafast optical pulse whose angle of   incidence is $\theta$.   The in-plane component of the incident pulse is oriented at an angle $\phi$.  (b) The first Brillouin zone of the reciprocal space of bilayer graphene. Points $K$ and $K^{'}$ are the Dirac points, which correspond to two valleys, 
and $\mathbf{b_{1}}$,$\mathbf{b_{2}}$ are the reciprocal lattice vectors.}
\label{fig1}
\end{center}
\end{figure}

\begin{figure}[ht!]
\begin{center}
\includegraphics[scale=0.5]{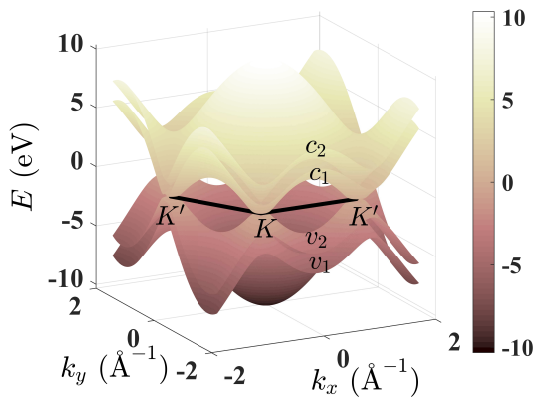}
\caption{ Energy dispersion  of bilayer graphene with four bands ($v_{1},v_{2},c_{1},c_{2}$). The Brillouin zone is shown by black line and the application of a perpendicular field of amplitude $F_{0}=0.5~\mbox{V}/\mbox{\AA}$ opens up a band-gap \cite{mccann2013} of $0.371$ eV between $v_{2}$ and $c_{1}$ at $K$ and $K^{\prime}$.}
\label{fig2}
\end{center}
\end{figure}

Monolayer of graphene has honeycomb crystal structure whose unit cell contains two non-equivalent carbon atoms A and B. The bilayer graphene, on the other hand, contains two coupled monolayers each of which has hexagonal lattice structure. The unit cell of bilayer graphene consists of four carbon atoms: with the contribution of two atoms each from the bottom and top layers. The bilayer graphene can have three configurations \cite{rozhkova2016}: AA stacking, AB stacking (Bernal stacking), and twisted bilayer. The AA stacking arises when each carbon atom in the top layer is exactly over the corresponding atom of the bottom layer. However, in AB-stacking configuration one of the carbon atom $B_{b}$ from the bottom layer  is placed exactly below the carbon atom $A_{t}$ of the top layer, as shown in Fig. \ref{fig1}(a). In the third configuration, the top graphene layer is rotated at some angle with respect to the bottom layer \cite{bistritzer2011}.  In this paper, we consider Bernal configuration of the bilayer graphene as shown in Fig. \ref{fig1}(a). The primitive reciprocal lattice vectors of the bilayer graphene are $\mathbf{b_{1}}=2\pi/a\left(1,1/\sqrt{3}\right)$, and $\mathbf{b_{2}}=2\pi/a\left(1,-1/\sqrt{3}\right)$.  The first Brillouin zone of the bilayer graphene is also a hexagon  whose vertices are the Dirac points at $K=2\pi/a\left(-1/3,1/\sqrt{3}\right)$ and $K^{\prime}=2\pi/a\left(1/3,1/\sqrt{3}\right)$, where $a=2.46~\mbox{\AA}$ is the lattice constant [Fig. \ref{fig1}(b)].  In AB-stacked bilayer graphene, the pair of sites (atom $B_{b}$ from the bottom layer and $A_{t}$ from the top layer), which exactly overlap, are referred as `dimers'. The interlayer coupling between the dimer sites is relatively strong because the orbitals of the dimer sites strongly overlap with each other. As a consequence of this, the hopping $\gamma_{1}$ between the dimer sites is the strongest. Thus, in such approximation we consider the following tight-binding Hamiltonian for the bilayer graphene with the nearest neighbor hopping \cite{rozhkova2016}
\begin{align}
H_{0}=\begin{bmatrix}
0 & -\gamma_{0} f\left(\mathbf{k}\right)& 0 & 0\\
-\gamma_{0} f^{\ast}\left(\mathbf{k}\right) & 0 & \gamma_{1} & 0\\
0 & \gamma_{1} & 0 & -\gamma_{0} f\left(\mathbf{k}\right)\\
0 & 0 & -\gamma_{0} f^{\ast}\left(\mathbf{k}\right) & 0
\end{bmatrix}\;,\label{eq1}
\end{align}
where $\gamma_{0}=3.16$ eV is the hopping integral, $\gamma_{1}=0.381$ eV represents the coupling between the orbitals on the dimer sites and $f(\mathbf{k})$ is given by the following expression
\begin{align}
f(\mathbf{k})=\exp\left(\frac{iak_{y}}{\sqrt{3}}\right)+2\exp\left(-\frac{iak_{y}}{2\sqrt{3}}\right)\cos\left(\frac{ak_{x}}{2}\right)\;.\label{eq2}
\end{align}
The energy spectrum of the bilayer graphene can be found from Eq. (\ref{eq1}). It consists of four bands: two conduction bands (CBs) and two valence bands (VBs) with the following energy dispersion 
\begin{align}
E_{c_{1}}&=-E_{v_{2}}=-\left(\gamma_{1}-\mathcal{F}\right)/2\;,\label{eq3} \\
E_{c_{2}}&=-E_{v_{1}}=\left(\gamma_{1}+\mathcal{F}\right)/2\;,\label{eq4}
\end{align}
where $\mathcal{F}=\sqrt{4\gamma_{0}^{2}|f(\mathbf{k})|^{2}+\gamma_{1}^{2}}$. The energy spectrum is shown in Fig. \ref{fig2}. 
The splitting between bands is determined by the interlayer coupling $\gamma_{1}$, as illustrated in Fig. \ref{fig2}(a) . At the Dirac points, two bands $c_{1}$ (CB) and $v_{2}$ (VB) are degenerate. The band structure of the bilayer graphene can be modified by applying a perpendicular electric field. This breaks the  interlayer symmetry and as a result the low-energy bands shows a `Mexican hat' shape with a band gap. 

\begin{figure}[ht!]
\begin{center}
\begin{tabular}{c}
\includegraphics[scale=0.5]{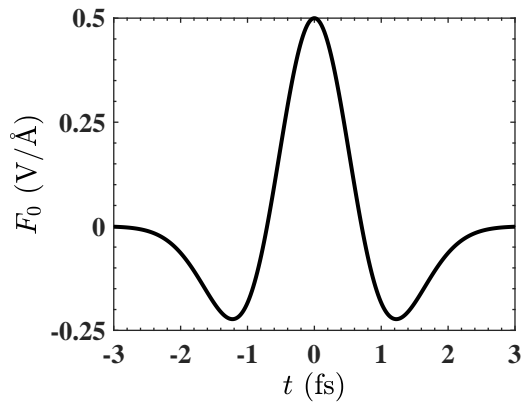}
\end{tabular}
\caption{The pulse profile as a function of time.  The parameters of the pulse are $F_{0}=0.5$ V/$\mbox{\AA}$ and $\tau=1$ fs.}
\label{fig3}
\end{center}
\end{figure}

Now, we consider a p-polarized ultrashort optical pulse of a single oscillation that is incident on bilayer graphene as shown in Fig. \ref{fig1}(a). The direction of 
propagation of the pulse is characterized by angle $\theta$, while the in-plane orientation of the pulse is determined by  angle $\phi$ measured with respect to the x-axis 
[see Fig. \ref{fig1}(a)]. In the presence of the optical pulse, the Hamiltonian of the system takes the following form
\begin{align}
H(t)=H_{0}-e\mathbf{F_{p}(t)}\mathbf{r}-\frac{eL_{z}F_{z}(t)}{2}\begin{bmatrix}
1 & 0& 0 & 0\\
0 & 1 & 0 & 0\\
0 & 0 & -1 & 0\\
0 & 0 & 0 & -1\end{bmatrix}\;,\label{eq5}
\end{align} 
where $e$ is the electron charge, $\mathbf{F_{p}(t)}=\left(F_{x}(t),F_{y}(t)\right)=F(t)\cos\theta(\cos\phi,\sin\phi)$ is the in-plane component of the electric field of the pulse, $L_{z}=3.35~\mbox{\AA}$ is the interlayer spacing, and $F_{z}(t)=F(t)\sin\theta$ represents the normal component. Note that here $F_{z}(t)$ creates the interlayer asymmetry by introducing difference in the on-site energies of the two layers and consequently it opens up a dynamic band gap at the $K$ and $K^{\prime}$ points. The applied ultrafast pulse is parametrized by the following equation 
\begin{align}
F&=F_{0}(1-2u^{2})e^{-u^{2}}\;,\label{eq6}
\end{align}
where $F_{0}$ is the amplitude of the optical pulse, and $u=t/\tau$, where $\tau$ is the pulse duration, which we consider to be 1 fs. The profile of the pulse in Eq. (\ref{eq6}) is displayed in Fig. \ref{fig3} and  has zero area, $\int_{-\infty}^{+\infty}F(t)dt=0$.

We assume that the electron dynamics in the presence of the external electric field of the optical pulse is coherent. This is because the duration of the pulse in Eq. (\ref{eq6}) is shorter than the characteristic electron scattering time of 10-100 fs \cite{breusing2011,malic2011,hwang2008}. The coherent electron dynamics is described by the time-dependent Schr\"{o}dinger equation
\begin{align}
i\hbar\frac{d\Psi}{dt}=H(t)\Psi\;.\label{eq7}
\end{align}
The applied strong electric field generates both interband and intraband electron dynamics. The intraband dynamics is determined by the Bloch acceleration theorem \cite{bloch1929}, which has following form:
\begin{align}
\hbar\frac{d\mathbf{k}}{dt}=e\mathbf{F_{p}(t)}\;.\label{eq8}
\end{align} 
Note that the acceleration theorem is universal and does not depend on the dispersion law. Thus, intraband electron dynamics is the same for all CBs and VBs. From Eq. (\ref{eq8}), the time-dependent crystal momentum, $\mathbf{k(\mathbf{q},t)}$, is written as
\begin{align}
\mathbf{k(\mathbf{q},t)}=\mathbf{q}+\frac{e}{\hbar}\int\limits_{-\infty}^{t}\mathbf{F_{p}}(t^{\prime})dt^{\prime}\;,\label{eq9}
\end{align}
where $\mathbf{q}$ is the initial crystal momentum. Since the intraband electron dynamics is universal in the reciprocal space, then the states which belong to different bands and have the same initial wave vector, $\mathbf{q}$, will have the same wave vector $\mathbf{k}(\mathbf{q},t)$ at later time $t$. For intraband electron dynamics the corresponding wave functions, which are the solutions of Eq. (\ref{eq7}) within a single band, i.e., without interband coupling, are well known Houston functions \cite{houston1940}
\begin{align}
\Phi_{\alpha q}^{(H)}(\mathbf{r},t)=\Psi^{(\alpha)}_{\mathbf{k}(\mathbf{q},t)}(\mathbf{r})\exp\left[i\left(\phi_{\alpha}^{(D)}(t)+\phi_{\alpha}^{(B)}(t)\right)\right]\;,\label{eq10}
\end{align}
where $\alpha=v_{1},v_{2},c_{1},c_{2}$ for VB and CB, respectively, and $\Psi_{\mathbf{k}}^{(\alpha)}$ are the periodic Bloch-band eigenfunctions in the absence of the optical pulse. Here $\phi_{\alpha}^{(D)}(t)$ is the dynamic phase in band $\alpha$,
\begin{align}
\phi_{\alpha}^{(D)}(t)&=-\frac{1}{\hbar}\int\limits_{-\infty}^{t} E_{\alpha}\left[\mathbf{k}(\mathbf{q},t^{\prime})\right]dt^{\prime}\;,\label{eq11}
\end{align}
and $\phi_{\alpha}^{(B)}(t)$ is the geometric  (Berry) phase in the band $\alpha$,
\begin{align}
\phi_{\alpha}^{(B)}(t)&=\frac{e}{\hbar}\int\limits_{-\infty}^{t}\mathbfcal{A}^{\alpha\alpha}\left(\mathbf{k}(\mathbf{q},t^{\prime})\right)\mathbf{F_{p}}(t^{\prime})dt^{\prime}\;,\label{eq12}
\end{align}
where the non-Abelian Berry connection \cite{wilczek1984} is given by
\begin{align}
\mathbfcal{A}^{\alpha\alpha^{\prime}}(\mathbf{q})=\Big\langle\Psi_{\mathbf{q}}^{(\alpha)}\vert i\frac{\partial}{\partial\mathbf{q}}\vert\Psi_{\mathbf{q}}^{(\alpha^{\prime})}\Big\rangle\;,\label{eq13}
\end{align}
where $\alpha,\alpha^{\prime}=v_{1},v_{2},c_{1},c_{2}$ and $\Psi_{\mathbf{q}}^{(\alpha)}$ are the periodic Bloch functions.  
Using the Houston functions (Eq. (\ref{eq10})) as the basis, we express the solution of Eq. (\ref{eq7}) in the following form
\begin{align}
\Psi_{\mathbf{q}(\mathbf{r},t)}=\sum_{\alpha=v_{1},v_{2},c_{1},c_{2}}\beta_{\alpha\mathbf{q}}(t)\Phi_{\alpha q}^{(H)}(\mathbf{r},t)\;,\label{eq14}
\end{align}
where $\beta_{\alpha\mathbf{q}}(t)$ are the expansion coefficients. 
%The coupling of VB to CB is determined by interband dipole matrix element defined as $\mathbf{D}^{\alpha_{i}\alpha_{j}}(\mathbf{q})=e\mathbf{\mathcal{A}^{\alpha_{i}\alpha_{j}}(\mathbf{q})}$. 

Note that the solution, Eq. (\ref{eq14}), is parametrized by initial electron wave vector $\mathbf{q}$ and the electron dynamics for the states with different values of $\mathbf{q}$ is decoupled. In the interaction picture the differential equations for the  expansion coefficients can be written as 
\begin{align}
i\hbar\frac{\partial B_{\mathbf{q}}(t)}{\partial t}&=H^{\prime}(\mathbf{q},t)B_{\mathbf{q}}(t)\;,\label{eq15}
\end{align}
where $B_{\mathbf{q}}(t)$, and  Hamiltonian $H^{\prime}(\mathbf{q},t)$ are defined as 
%\begin{widetext}
\begin{align}
B_{\mathbf{q}}(t)&=\begin{bmatrix}
\beta_{c_{2}\mathbf{q}}(t)\\ \beta_{c_{1}\mathbf{q}}(t)\\ \beta_{v_{1}\mathbf{q}}(t)\\ \beta_{v_{2}\mathbf{q}}(t)
\end{bmatrix}\;,\label{eq16}\\
H^{\prime}(\mathbf{q},t)&=-e\mathbf{F}(t)\hat{\mathbfcal{A}_{3d}}(\mathbf{q},t)\;,\label{eq17}\\
\hat{\mathbfcal{A}_{3d}}(\mathbf{q},t)&=\nonumber\\
&\begin{bmatrix}
0 & \mathbfcal{D}^{c_{2}c_{1}}_{3d}(\mathbf{q},t) & \mathbfcal{D}^{c_{2}v_{1}}_{3d}(\mathbf{q},t) & \mathbfcal{D}^{c_{2}v_{2}}_{3d}(\mathbf{q},t)\\
\mathbfcal{D}^{{c_{1}c_{2}}}_{3d}(\mathbf{q},t)& 0  & \mathbfcal{D}^{c_{1}v_{1}}_{3d}(\mathbf{q},t) & \mathbfcal{D}^{c_{1}v_{2}}_{3d}(\mathbf{q},t)\\
\mathbfcal{D}^{{v_1}c_{2}}_{3d}(\mathbf{q},t) & \mathbfcal{D}^{{v_{1}c_{1}}}_{3d}(\mathbf{q},t) & 0 & \mathbfcal{D}^{v_{1}v_{2}}_{3d}(\mathbf{q},t)\\
\mathbfcal{D}^{{v_{2}c_{2}}}_{3d}(\mathbf{q},t) & \mathbfcal{D}^{{v_{2}c_{1}}}_{3d}(\mathbf{q},t) & \mathbfcal{D}^{{v_{2}v_{1}}}_{3d}(\mathbf{q},t) & 0
\end{bmatrix}\;,\label{eq18}
\end{align}
%\end{widetext}
where $\mathbf{F}(t)=\mathbf{F}_{\mathbf{p}}(t)+F_{z}(t)$, and $\mathbfcal{D}^{\alpha^{\prime}\alpha}_{3d}=(\mathbfcal{D}^{\alpha\alpha^{\prime}}_{3d})^{\ast}$. Also we have introduced following quantities 
\begin{align}
\mathbfcal{D}_{3d}^{\alpha\alpha^{\prime}}(\mathbf{q},t)&=\mathbfcal{D}_{\mathbf{p}}^{\alpha\alpha^{\prime}}(\mathbf{q},t)+\mathcal{D}^{\alpha\alpha^{\prime}}_{z}(\mathbf{q},t)\nonumber\\
&=\Big(\mathbfcal{A}^{\alpha\alpha^{\prime}}[\mathbf{k}(\mathbf{q},t)]+\mathcal{A}^{\alpha\alpha^{\prime}}_{z}[\mathbf{k}(\mathbf{q},t)]\Big)\nonumber\\
&\times\exp\Big(i(\phi_{\alpha\alpha^{\prime}}^{(D)}(\mathbf{q},t)+\phi_{\alpha\alpha^{\prime}}^{(B)}(\mathbf{q},t))\Big)\;,\label{eq19}\\
\phi_{\alpha\alpha^{\prime}}^{(D)}(\mathbf{q},t)&=\phi_{\alpha^{\prime}}^{(D)}(\mathbf{q},t)-\phi_{\alpha}^{(D)}(\mathbf{q},t)\;,\label{eq20}\\
\phi_{\alpha\alpha^{\prime}}^{(B)}(\mathbf{q},t)&=\phi_{\alpha^{\prime}}^{(B)}(\mathbf{q},t)-\phi_{\alpha}^{(B)}(\mathbf{q},t)\;.\label{eq21}
\end{align}
Here $\mathbfcal{D}_{\mathbf{p}}^{\alpha\alpha^{\prime}}(\mathbf{q},t)$ and $\mathcal{D}^{\alpha\alpha^{\prime}}_{z}(\mathbf{q},t)$ represents the respective in-plane and normal component of the non-Abelian Berry connection. Specifically, $\mathcal{D}^{\alpha\alpha^{\prime}}_{z}(\mathbf{q},t)$ represents the interband coupling introduced by the normal component of the electric field of the applied laser pulse. The analytic expressions of $\mathbfcal{A}^{\alpha\alpha^{\prime}}$, and $\mathcal{A}_{z}^{\alpha\alpha^{\prime}}$ are provided in the Appendix \ref{appendixA}.

The time-dependent electric field of the optical pulse causes the polarization of the system which generates an electric current.  Both intraband ($\mathbf{J}^{intra}(t)$) and interband ($\mathbf{J}^{inter}(t)$) currents contribute to the total current, $\mathbf{J}(t)=\mathbf{J}^{intra}(t)+\mathbf{J}^{inter}(t)$, in the system. Here, intraband and interband currents can be expressed in the following form:
\begin{align}
\mathbf{J}^{intra}(t)&=2e\int d\mathbf{q}\sum_{\alpha=v_{1},v_{2},c_{1},c_{2}}\beta^{\ast}_{\alpha\mathbf{q}}V_{j}^{\alpha}\beta_{\alpha\mathbf{q}}\;,\label{eq22}\\
\mathbf{J}^{inter}(t)&=2e\int d\mathbf{q}\underset{\alpha\neq\alpha^{\prime}}{\sum_{\alpha,\alpha^{\prime}=v_{1},v_{2},c_{1},c_{2}}}\beta^{\ast}_{\alpha\mathbf{q}}V_{j}^{\alpha\alpha^{\prime}}\beta_{\alpha^{\prime}\mathbf{q}}\;,\label{eq23}
\end{align}
where $V_{j}^{\alpha\alpha^{\prime}}$ is the interband velocity which are the matrix elements of the velocity operator $\hat{V}_{j}=\frac{1}{\hbar}\frac{\partial H_{0}}{\partial k_{j}}$, $V_{j}^{\alpha}$ is the intraband velocity (group velocity) and  $j=x,y$. Here factor of 2 is due to spin degeneracy. The interband velocities can be expressed in terms of the non-Abelian Berry connection 
\begin{align}
V_{j}^{\alpha\alpha^{\prime}}=\frac{i}{\hbar}\mathcal{A}_{j}^{\alpha\alpha^{\prime}}\left(E_{\alpha}-E_{\alpha^{\prime}}\right)\;.\label{eq24}
\end{align} 
The matrix elements of the intraband velocity operator are presented in Appendix \ref{appendixB}.

\section{Results}
\label{results}
By using the formalism described in Sec. \ref{model}, we numerically solve the system of Eqs. (\ref{eq15}) under the initial conditions of initially occupied valence bands 1 and 2 i.e. $(\beta_{c_{2}\mathbf{q}},\beta_{c_{1}\mathbf{q}},\beta_{v_{1}\mathbf{q}},\beta_{v_{2}\mathbf{q}})=(0,0,1,0)$ and $(\beta_{c_{2}\mathbf{q}},\beta_{c_{1}\mathbf{q}},\beta_{v_{1}\mathbf{q}},\beta_{v_{2}\mathbf{q}})=(0,0,0,1)$ . From the solution of Eqs. (\ref{eq15}), we calculate the CB populations after the pulse which is known as residual CB populations, $N_{c_{2}}^{(\mbox{res})}(\mathbf{q})=|\beta_{c_{2}\mathbf{q}}(t=\infty)|^{2}$ and $N_{c_{1}}^{(\mbox{res})}(\mathbf{q})=|\beta_{c_{1}\mathbf{q}}(t=\infty)|^{2}$. To further characterize the interband dynamics we define the time-dependent total population of the conduction bands
\begin{align}
\mathcal{N}_{CB,\alpha}(t)=\sum\limits_{\mathbf{q},i}|\beta_{\alpha,\mathbf{q}}^{(i)}(t)|^{2}\;,\label{eq25}
\end{align}
where $\alpha=c_{1},c_{2}$ and $i=1,2$ corresponds to initial conditions described above.

\subsection{x-polarized pulse}

\begin{figure}[ht!]
\begin{center}
\begin{tabular}{c}
\includegraphics[scale=0.27]{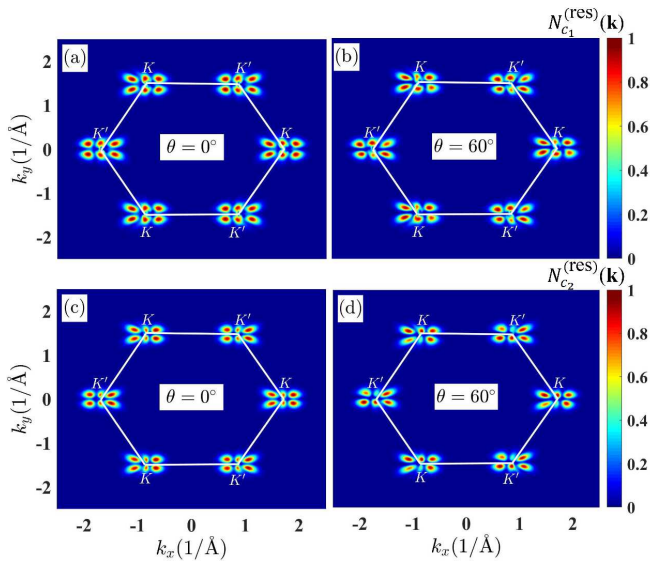}
\end{tabular}
\end{center}
\caption{Residual CB population (a,b) $N^{\mbox{(res)}}_{c_{1}}(\mathbf{k})$ and (c,d) $N^{\mbox{(res)}}_{c_{2}}(\mathbf{k})$ of two CBs for bilayer graphene in the reciprocal space as a function of the wave vector ($\mathbf{k}$) for different angles of incidence ($\theta$). The boundary of the first Brillouin zone is shown by white lines. The amplitude of the pulse is fixed at $F_{0}=0.5$ V/$\mbox{\AA}$, while the angle of incidence is (a,c) $\theta=0^{\circ}$, and (b,d) $\theta=60^{\circ}$. Here $\phi=0^{\circ}$, and rest of the parameters are same as in Fig. \ref{fig3}.}
\label{fig4}
\end{figure}

The electron dynamics in bilayer graphene depends on the angle of incidence and the intensity of the applied pulse. To explore this effect,  we first consider the case of $\phi=0^{\circ}$, i.e., when the pulse is polarized along the $x$-direction. Also, to probe the influence of the oblique incidence, we fix the incidence angle dependence in the plane of the bilayer graphene. Thus, the in-plane field amplitude of the pulse remains fixed. In Fig. \ref{fig4}, we show the distribution of  the residual CB population in the first Brillouin zone  at the end of the linearly polarized pulse for different angles of incidence and for fixed amplitude of the pulse,  0.5 V/$\mbox{\AA}$. For the normally incident pulse, see Fig. \ref{fig4}(a,c), the CB population distributions is symmetric with respect to the $x$-axis and have hot spots with large CB population separated by dark regions with small CB population. These distributions are similar to the one that was observed in monolayer graphene \cite{kelardeh2015} and are due to interference that is caused by the double passage by electrons of the Dirac points during one cycle of the optical pulse. 

At a finite angle of incidence, the normal component of the optical field opens up a band gap between valence band 2 and conduction band 1. This will cause an asymmetry in CB population distribution with respect to the $x$-axis, which is clearly visible in Fig. \ref{fig4}(b,d). Further, the CB population distribution can be reversed for the negative angle of incidence. 

Redistribution of electrons between the valence and conduction bands results in the generation of electric currents during and after the pulse, governed by  Eqs. (\ref{eq22}) and (\ref{eq23}). Here we explore the dependence of the ultrafast currents the angle of incidence and the amplitude of the linearly polarized pulse. 
\begin{figure}[ht!]
\begin{center}
\begin{tabular}{cc}
\includegraphics[scale=0.5]{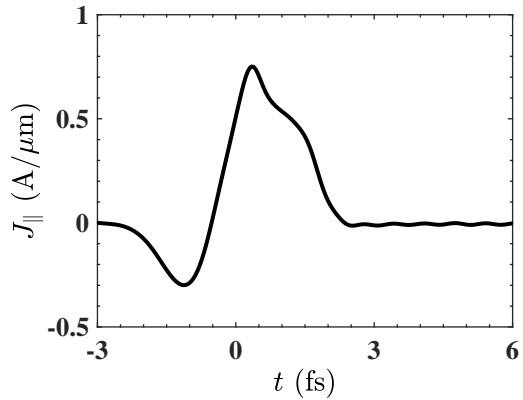}
\end{tabular}
\end{center}
%\vspace*{-1cm}
\caption{(a) Electric current density along the longitudinal direction of the polarization of the pulse as a function of time. Here $\theta=0^{\circ}$, $\phi=0^{\circ}$ and $F_{0}=0.5~\mbox{V}/\mbox{\AA}$. The other parameters are same as in Fig. \ref{fig4}.}
\label{fig5}
\end{figure}

Since the CB population distribution is symmetric with respect to the $x$-axis for $\theta =0$, then the corresponding generated electric current is longitudinal i.e. it has only $x$-component, see Fig. \ref{fig5},  while the transverse component is zero. Both interband and intraband electron dynamics produce the current in the system but the net electric current is mainly  determined by the intraband contribution. However, in long time run such a longitudinal current shows small oscillations [see Fig. \ref{fig5}] which are contributed by interband currents.

\begin{figure}[ht!]
\begin{center}
\begin{tabular}{cc}
\includegraphics[scale=0.5]{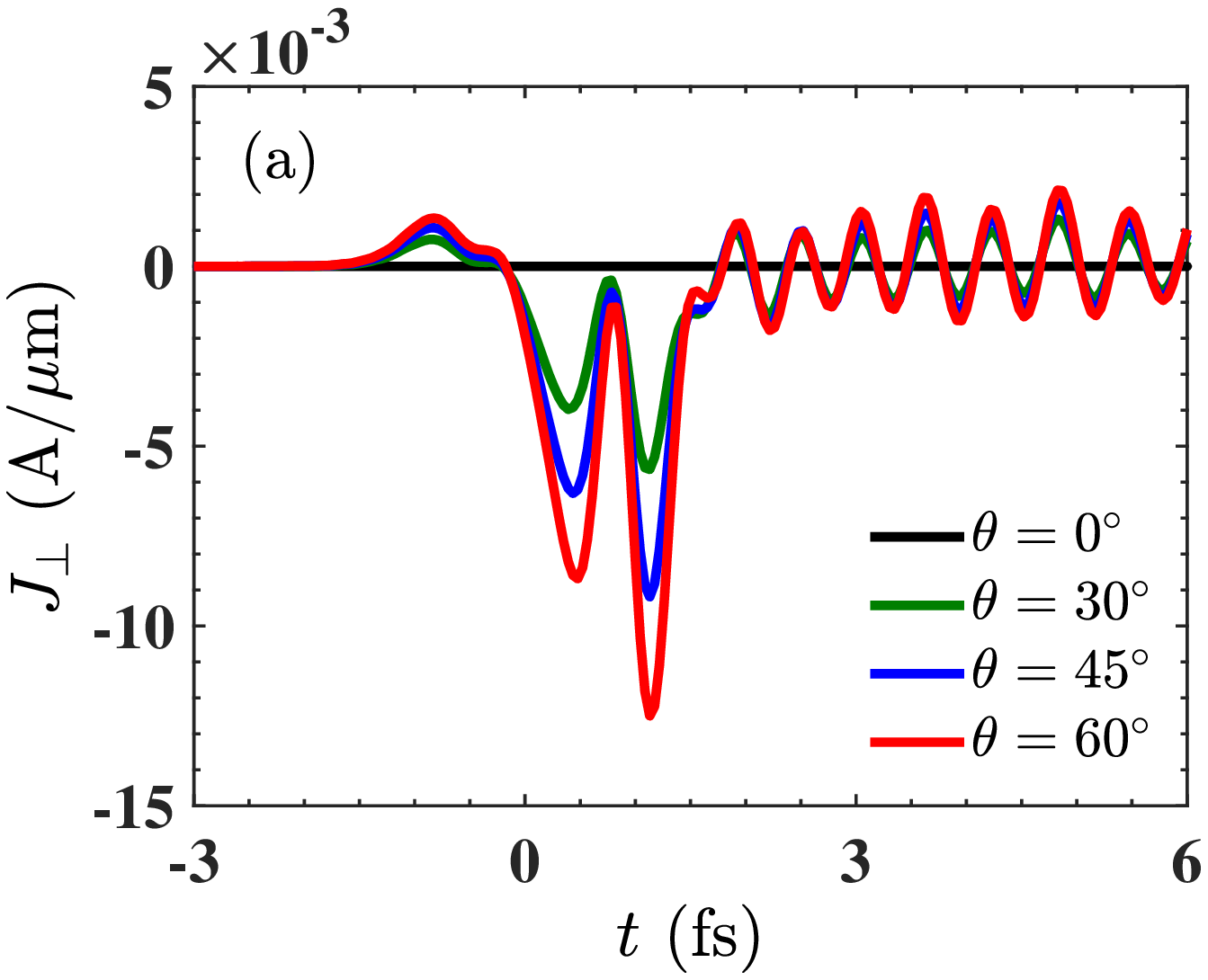}\\
\includegraphics[scale=0.5]{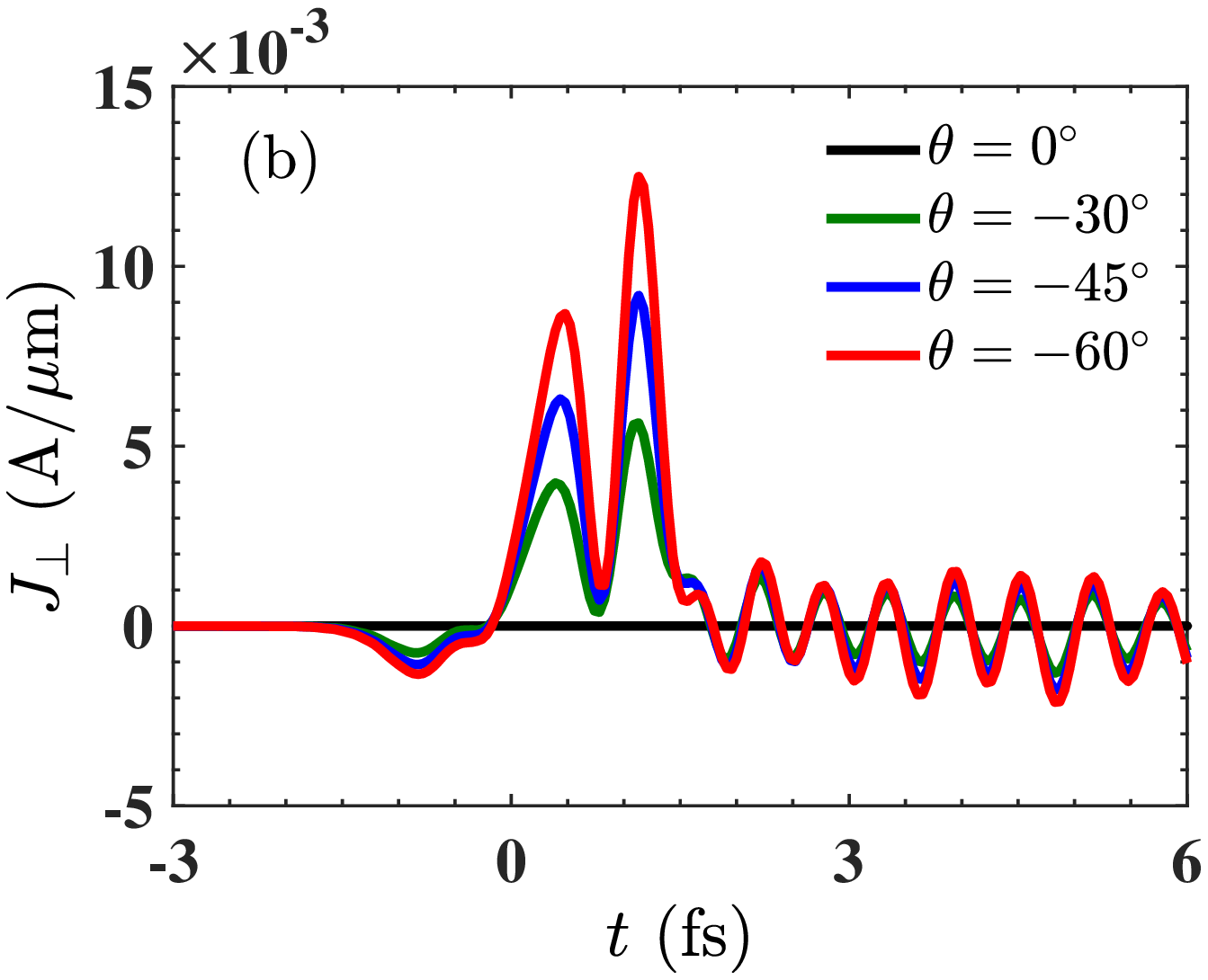}
\end{tabular}
\end{center}
%\vspace*{-1cm}
\caption{(a) Electric current density along the direction transverse to the polarization of the optical pulse for different angles of incidence. (b) The same as (a) except for the negative angles of incidence. The parameters are same as in Fig. \ref{fig5}.}
\label{fig6}
\end{figure}

Note that here the polarization of the pulse is along $x-$direction which is not the axis of symmetry of the bilayer graphene and for the oblique incidence of the ultrafast pulse, the normal component of the applied electric field produces interlayer asymmetry, which opens up a band gap between the CB and VB. As a consequence the residual CB population distribution is asymmetric with respect to the $x$-axis due to which there is a generation of a transverse current i.e. along the $y$-direction if the pulse is polarized along $x$-direction as shown in  Fig. \ref{fig6}(a). Such a behavior is different from graphene where no transverse current is predicted if the polarization of the pulse is along  $x$-axis which also corresponds to its  axis of symmetry. Further, the transverse current increases with increase in the angle of incidence. For negative angles of incidence the electron dynamics along the direction perpendicular to the polarization of the pulse gets reversed due to which the transverse electric current changes sign as shown in  Fig. \ref{fig6}(b).

\begin{figure}[ht!]
\begin{center}
\begin{tabular}{cc}
\includegraphics[scale=0.5]{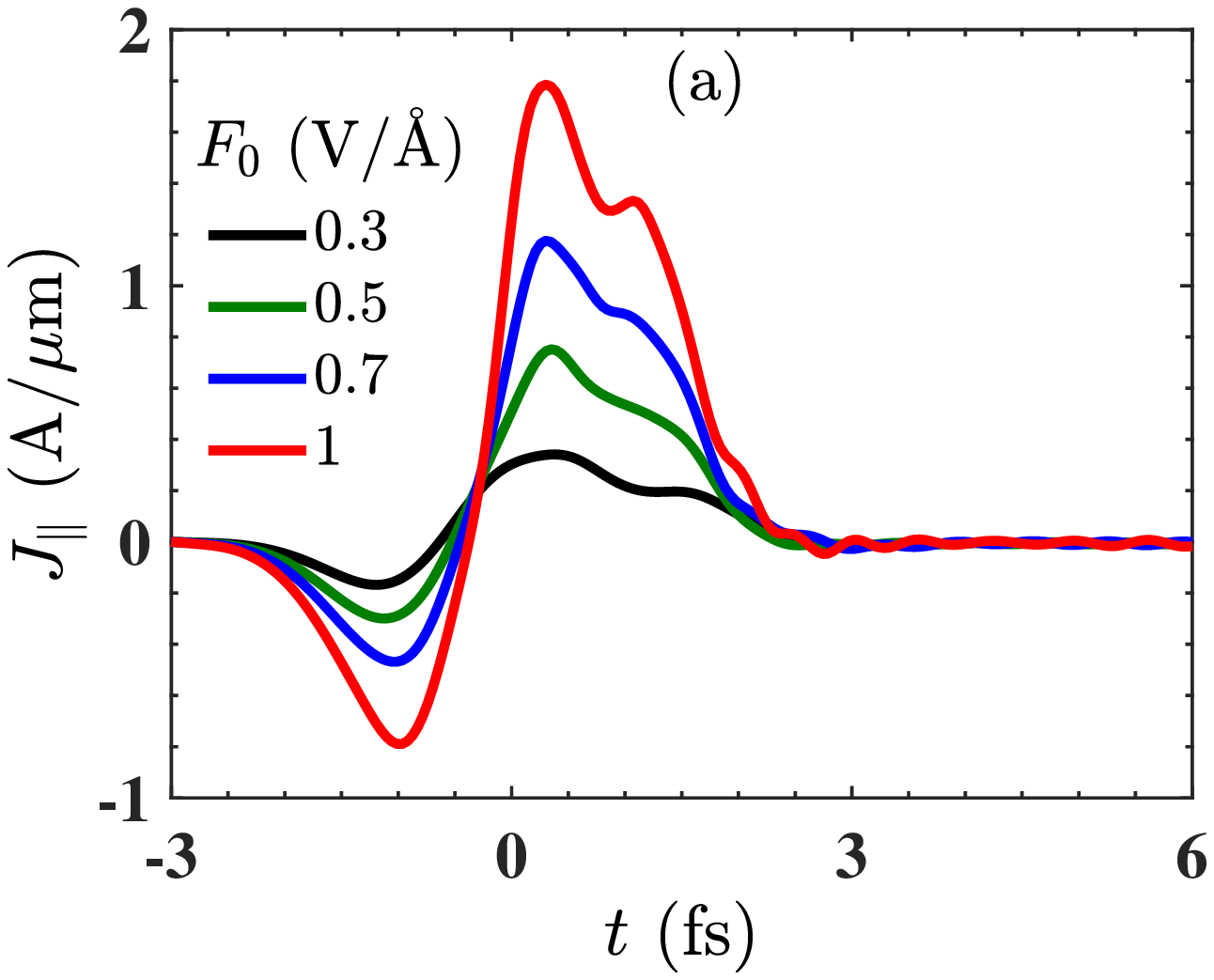}\\
\includegraphics[scale=0.5]{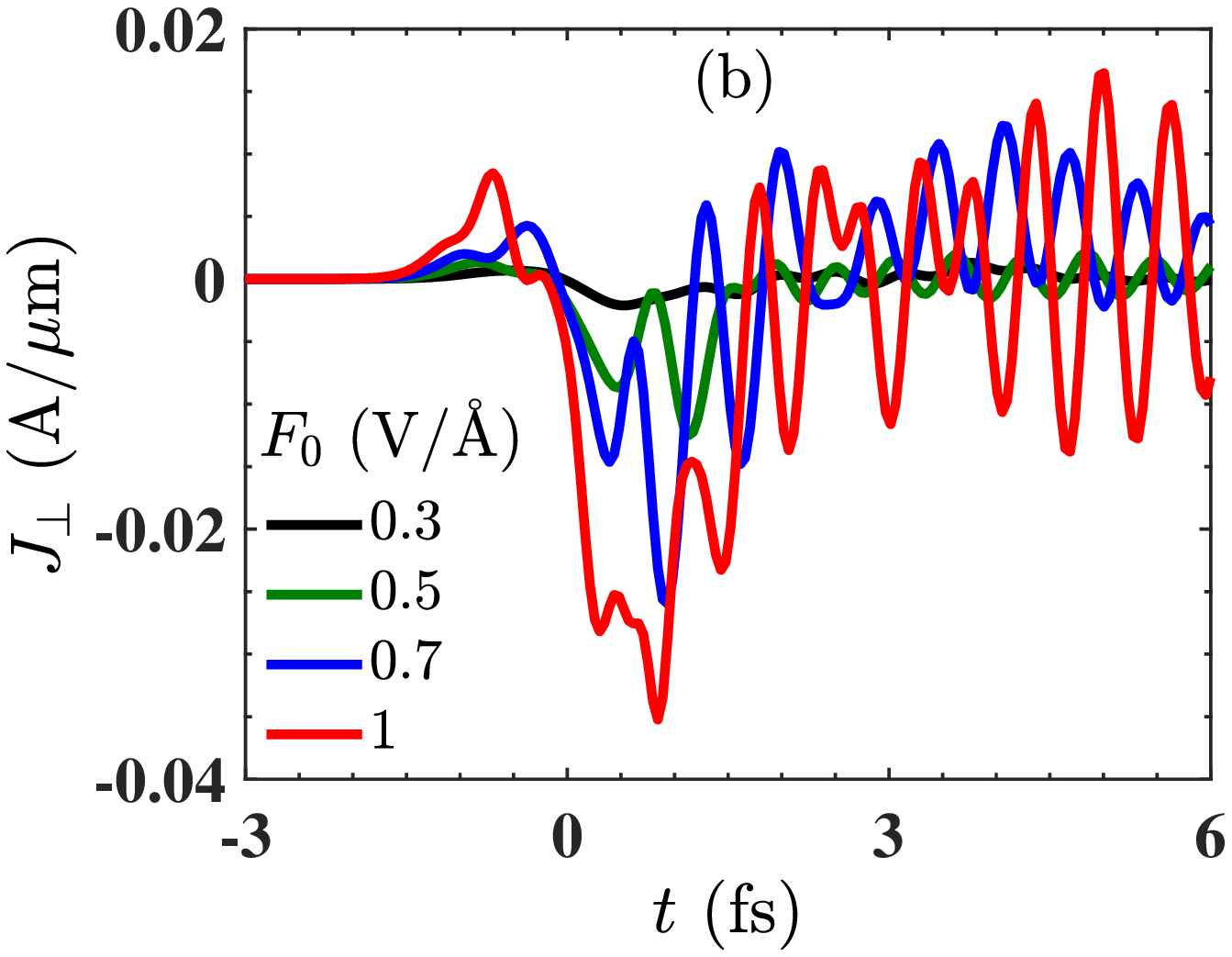}
\end{tabular}
\end{center}
%\vspace*{-1cm}
\caption{Electric current density along the direction (a) longitudinal and (b) transverse to the polarization of the optical pulse for different field amplitudes.  Here $\theta=60^{\circ}$ and rest of the parameters are same as in Fig. \ref{fig5}.}
\label{fig7}
\end{figure}

The electron redistribution between CBs and VBs also depends on the amplitude of the pulse. This influences the longitudinal as well as transverse current and the magnitude of the both increases with increase in field amplitude, as shown in Fig. \ref{fig7}. However, the longitudinal current switches sign if the direction of the field maximum is reversed whereas the transverse current remains unaffected.

The current generated in the system results in the  charge transfer through the system, which can be calculated from the following expression  
\begin{align}
Q_{j}=\int_{-\infty}^{+\infty}dt J_{j}(t)e^{-t/\tau_{r}}\;,\label{eq26}
\end{align}
where $j=\parallel,\perp$. Note that in Eq. (\ref{eq26}), we introduce the exponential decay in current with a relaxation time $\tau_{r}$.  This is done to relax the oscillations in the current with increase in the integration time. Practically it is relevant since increase in integration time results in the introduction of relaxation in the electron dynamics and consequently decay in the current. Here for numerical calculations, we set $\tau_{r}=10$ fs.

\begin{figure}[ht!]
\begin{center}
\begin{tabular}{cc}
\includegraphics[scale=0.5]{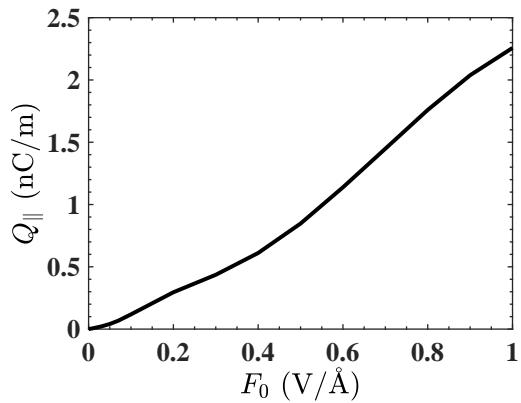}
\end{tabular}
\end{center}
%\vspace*{-1cm}
\caption{The transferred charge density in the direction parallel to the $x-$polarized optical pulse as a function of time. The other parameters are same as in Fig. \ref{fig5}.}
\label{fig8}
\end{figure}

The charge transferred  along the longitudinal direction is shown in Fig. \ref{fig8} as a function of the field amplitude. Two main features of the charge transfer are to be noted from Fig. \ref{fig8}: (i) The magnitude of the charge transfer increases with increase in the field amplitude. (ii) The direction of charge transfer is positive which means that it follows the direction of field maximum. 

\begin{figure}[ht!]
\begin{center}
\begin{tabular}{cc}
\includegraphics[scale=0.5]{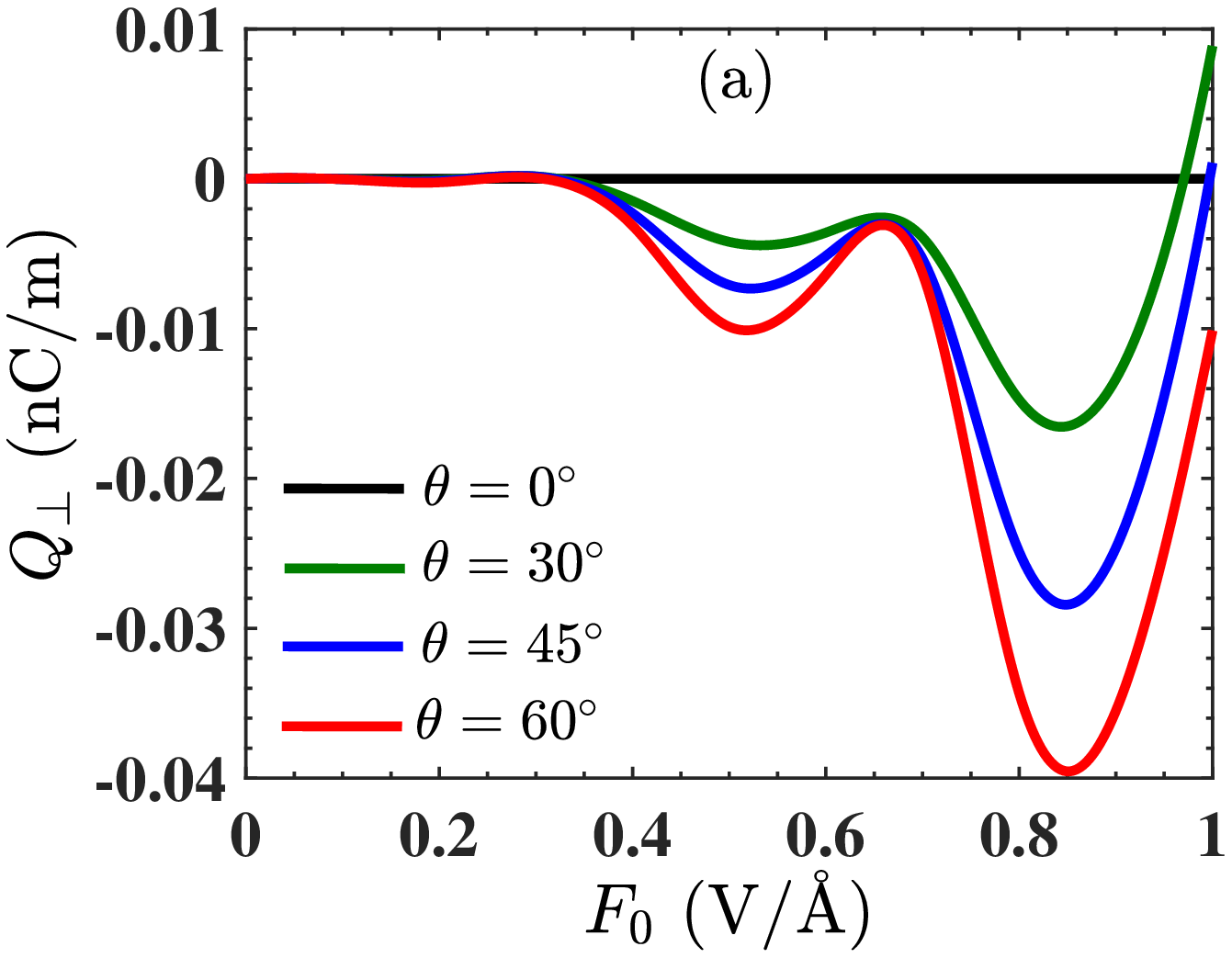}\\
\includegraphics[scale=0.5]{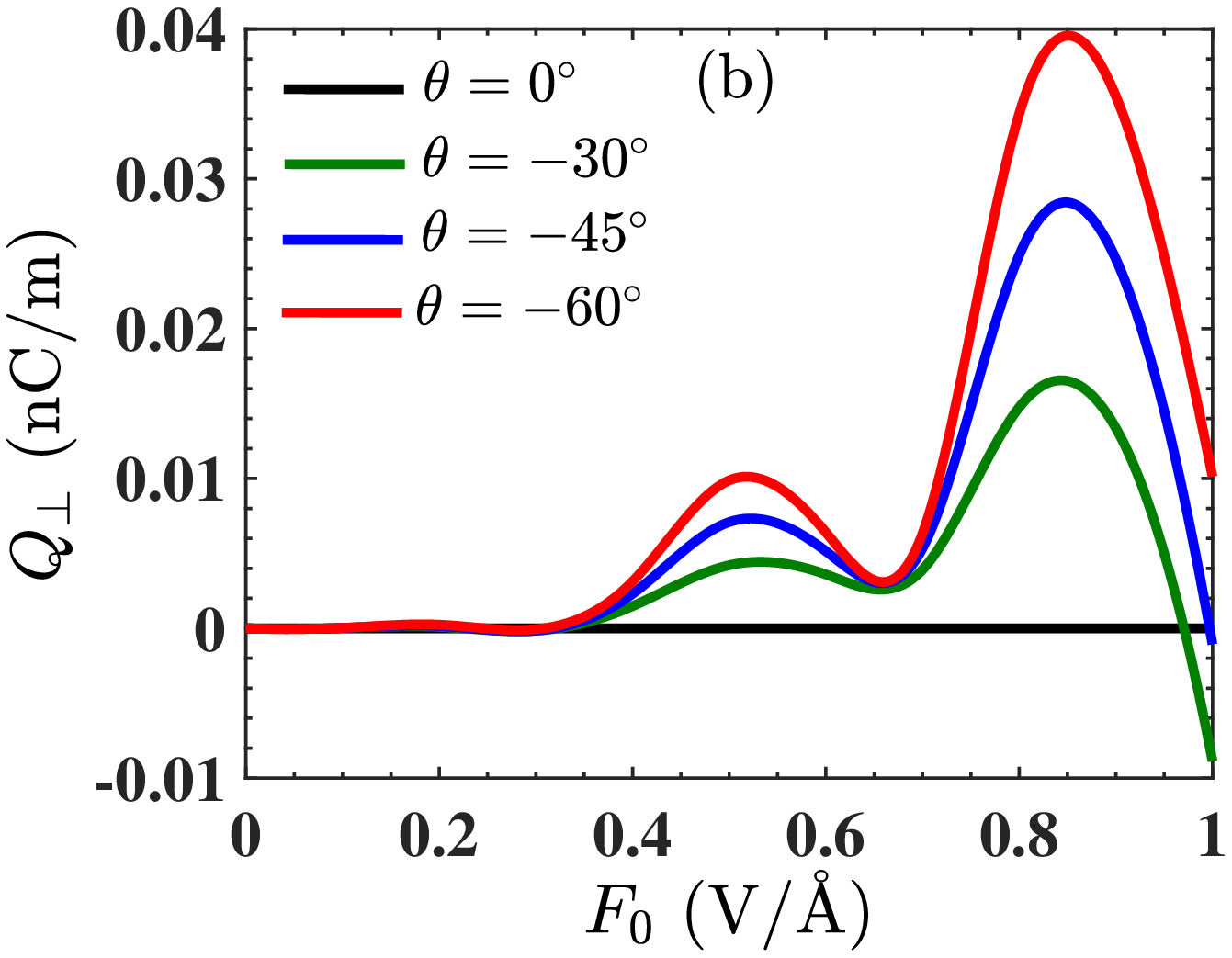}
\end{tabular}
\end{center}
%\vspace*{-1cm}
\caption{(a) The charge density transferred along a  direction transverse to the plane of incidence of the optical pulse as a function of the field amplitude for different 
angles of incidence. (b) The same as (a) except for negative angles of incidence. The parameters are same as in Fig. \ref{fig5}.}
\label{fig9}
\end{figure}

In Fig. \ref{fig9}, we show the charge transferred along the transverse direction i.e. along the $y$-direction for different angles of incidence. The transferred charge increases with the field amplitude and with the angle of incidence. The direction of the charge transfer depends on the sign of the $z$ component of the field maximum. Namely, for positive angle $\theta$, i.e., for the $z$-component of the field maximum is positive, the charge is transferred in the negative direction of the $z$-axis, while for negative angle $\theta$, which corresponds to the negative $z$-component of the field maximum, the transferred charge is positive. However, at $F_{0}=1~\mbox{V}/\mbox{\AA}$ and $\theta>\pm 44^{\circ}$, the charge transfer switches sign both for positive and negative angles of incidences, as shown in Fig. \ref{fig9}.

\begin{figure}[ht!]
\begin{center}
\begin{tabular}{cc}
\includegraphics[scale=0.5]{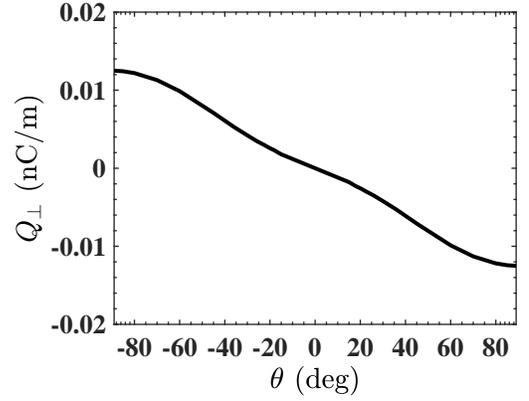}
\end{tabular}
\end{center}
%\vspace*{-1cm}
\caption{Transferred charge along the direction transverse to the plane of incidence of the optical pulse as a function of the angle of incidence for the field amplitude of $F_{0}=0.5~\mbox{V}/\mbox{\AA}$. Rest of the parameters are same as in Fig. \ref{fig5}.}
\label{fig10}
\end{figure}

The dependence of the charge transferred in the transverse direction, $Q_{\perp}$, on the angle of incidence is shown in Fig. \ref{fig10}. The transferred charge has almost linear dependence on angle $\theta$. It is positive for negative $\theta$ and odd function of the angle. 
\subsection{y-polarized pulse}
\begin{figure}[ht!]
\begin{center}
\begin{tabular}{c}
\includegraphics[scale=0.28]{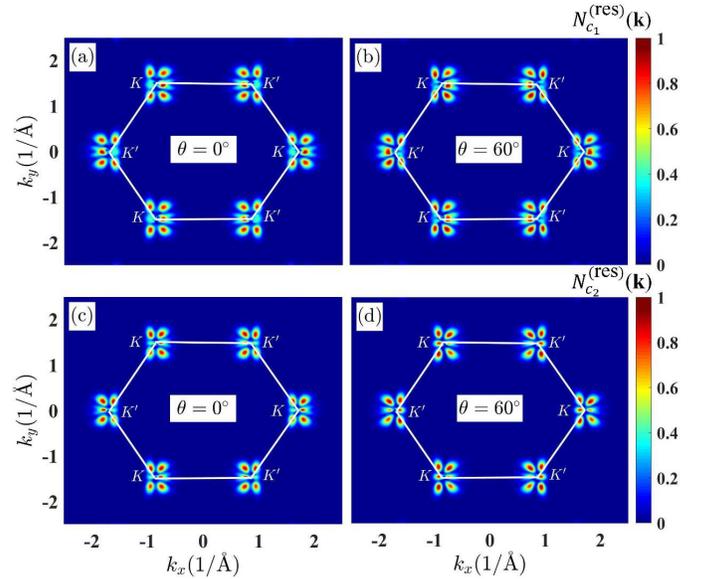}
\end{tabular}
\end{center}
\caption{Residual CB population (a,b) $N^{\mbox{(res)}}_{c_{1}}(\mathbf{k})$ and (c,d) $N^{\mbox{(res)}}_{c_{2}}(\mathbf{k})$ of two CBs in bilayer graphene for $y-$polarized pulse in the reciprocal space as a function of the wave vector ($\mathbf{k}$) for different angles of incidence ($\theta$). The boundary of the first Brillouin zone is shown by white lines. The amplitude of the pulse is fixed at $F_{0}=0.5$ V/$\mbox{\AA}$, while the angle of incidence is (a,b) $\theta=0^{\circ}$, and (c,d) $\theta=60^{\circ}$. Here $\phi=90^{\circ}$, and rest of the parameters are same as in Fig. \ref{fig3}.}
\label{fig11}
\end{figure}

In the preceding section, the pulse was polarized in the $x$-$z$ plane and the electric current was generated along the $y$-direction due to the braking of the interlayer symmetry by the normal component of the applied pulse. Now we fix $\phi=90^{\circ}$ i.e. the polarization of pulse is along $y-$axis which is also the axis of symmetry of bilayer graphene. For this case, no symmetry braking occurs and the CB distribution along y-axis remains partially symmetric even for non-zero angle of incidence as shown in Fig. \ref{fig11}. 

\begin{figure}[ht!]
\begin{center}
\begin{tabular}{cc}
\includegraphics[scale=0.5]{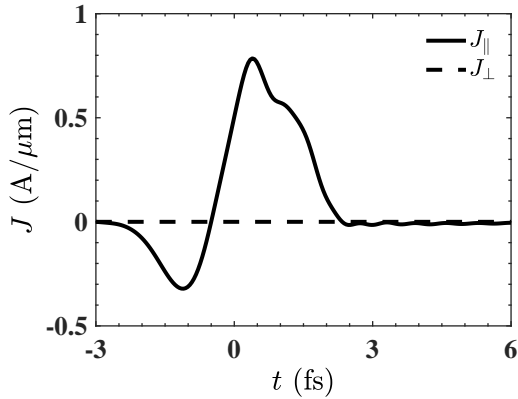}
\end{tabular}
\end{center}
%\vspace*{-1cm}
\caption{Electric current density parallel (solid black line) and perpendicular (dashed line) to the $y-$polarization of the optical pulse as a function of time. Here $\theta=60^{\circ}$ and rest of the parameters are same as in Fig. \ref{fig5}.}
\label{fig12}
\end{figure}

The current produced from  such a distribution is shown in Fig. \ref{fig12} for an angle of incidence of 60$^{\circ}$ and field amplitude of 0.5 $\mbox{V}/\mbox{\AA}$. There is a longitudinal current i.e. the current along y-direction but the current along the transverse direction i.e. along the $x-$axis remains zero during the pulse. This is attributed to the fact that here the pulse is applied along the axis of symmetry which forbids the generation of transverse current. 
\subsection{CB population dynamics}
\begin{figure}[ht!]
\begin{center}
\begin{tabular}{cc}
\includegraphics[scale=0.5]{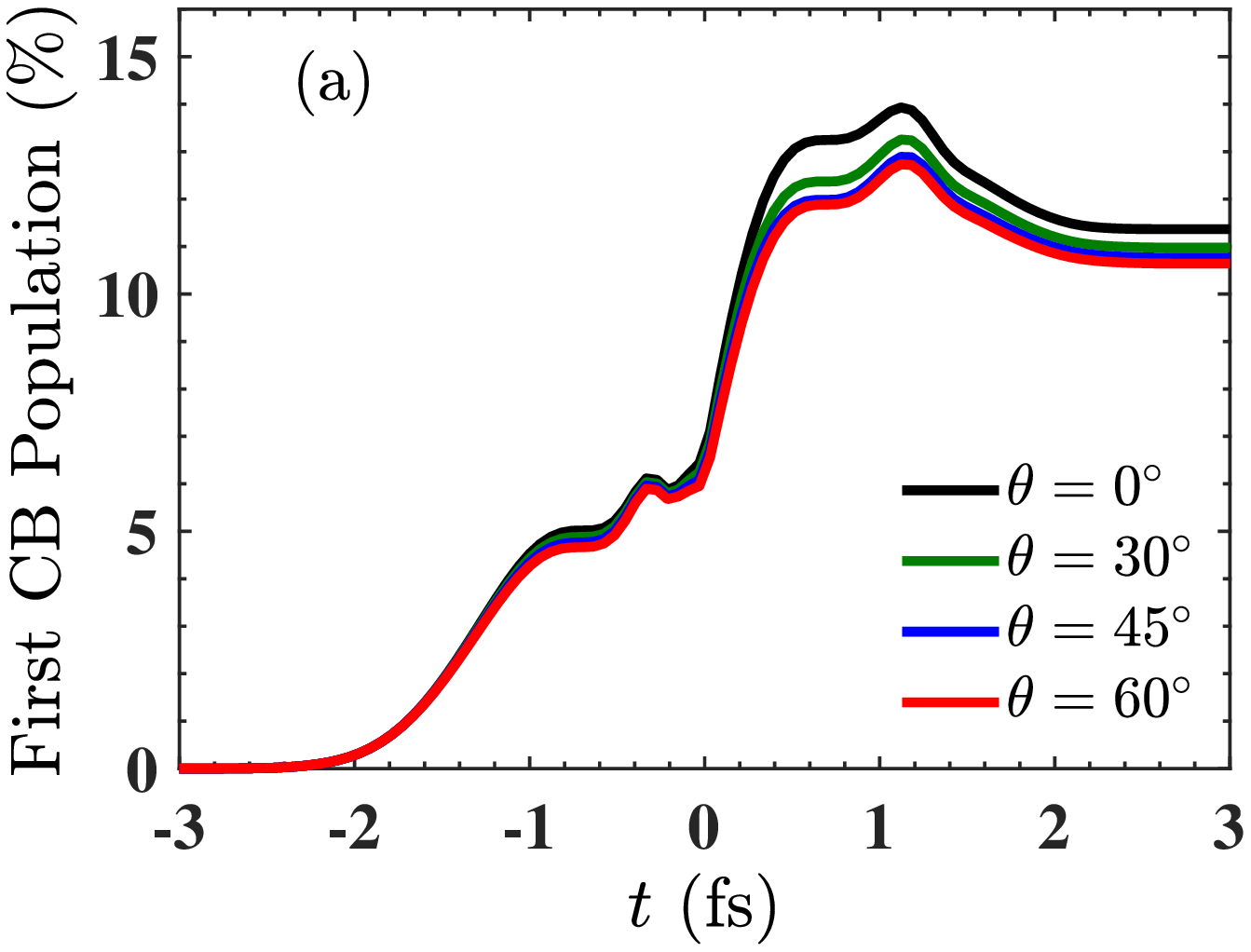}\\
\includegraphics[scale=0.5]{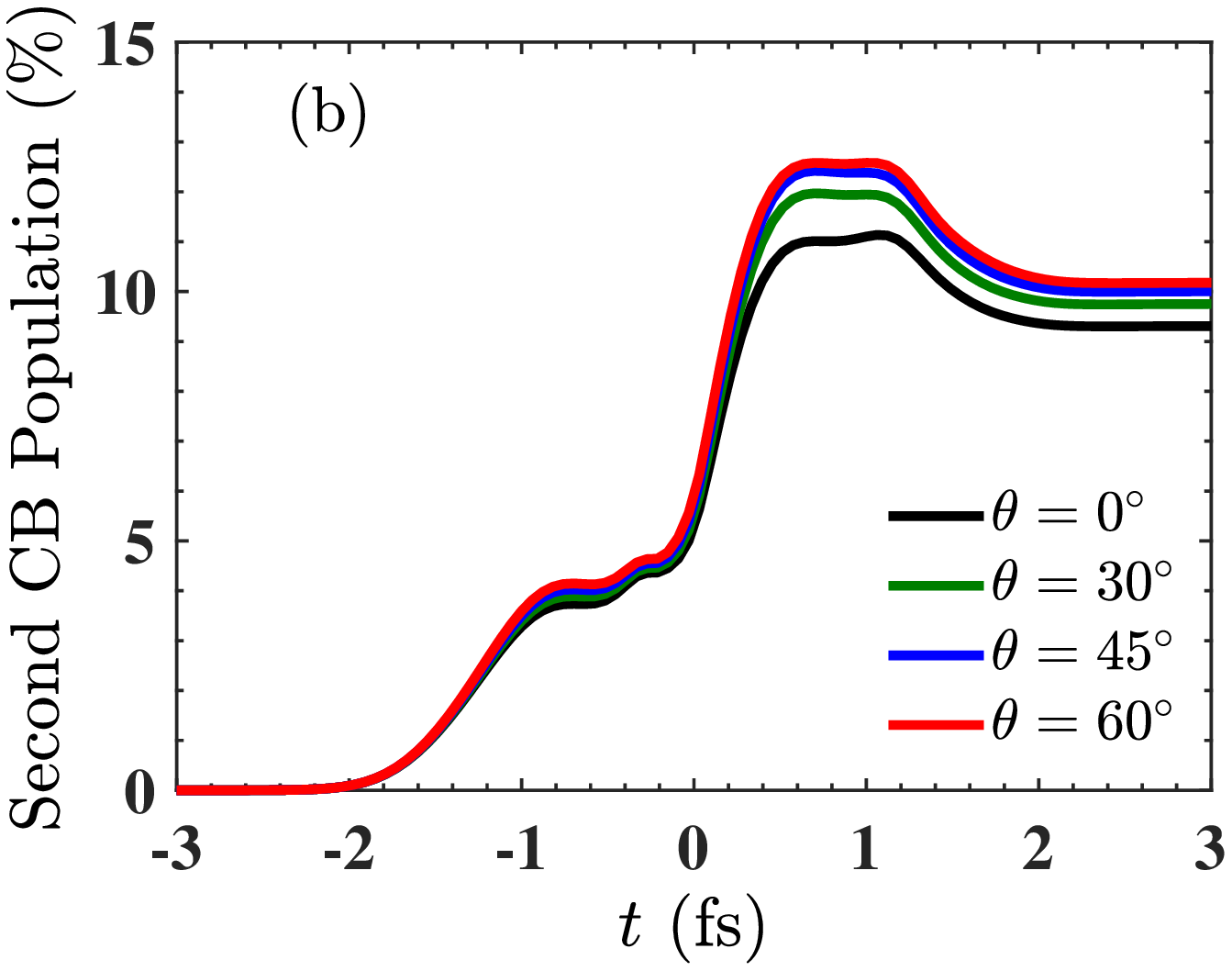}
\end{tabular}
\end{center}
%\vspace*{-1cm}
\caption{Population of (a) the first CB and (b) the second CB as a function of time for different angles of incidence of the optical pulse. 
The other parameters are the same as in Fig. \ref{fig5}.}
\label{fig13}
\end{figure}

One of the characteristics of the interband electron dynamics is the temporal evolution of the total CB population given by Eq. (\ref{eq25}). Such time dependence of the total population of the first and the second CBs are shown in Fig. \ref{fig13} for different values of the angle of incidence. The data show that the electron dynamics is irreversible and the residual CB populations for both bands are comparable to the maximum populations during the pulse. The dependence of the CB populations on the angle of incidence is different for two bands, while the CB population of the first CB decreases with $\theta $, the CB population of the second CB increases with $\theta $, see Fig. \ref{fig13}. 

\begin{figure}[ht!]
\begin{center}
\begin{tabular}{cc}
\includegraphics[scale=0.5]{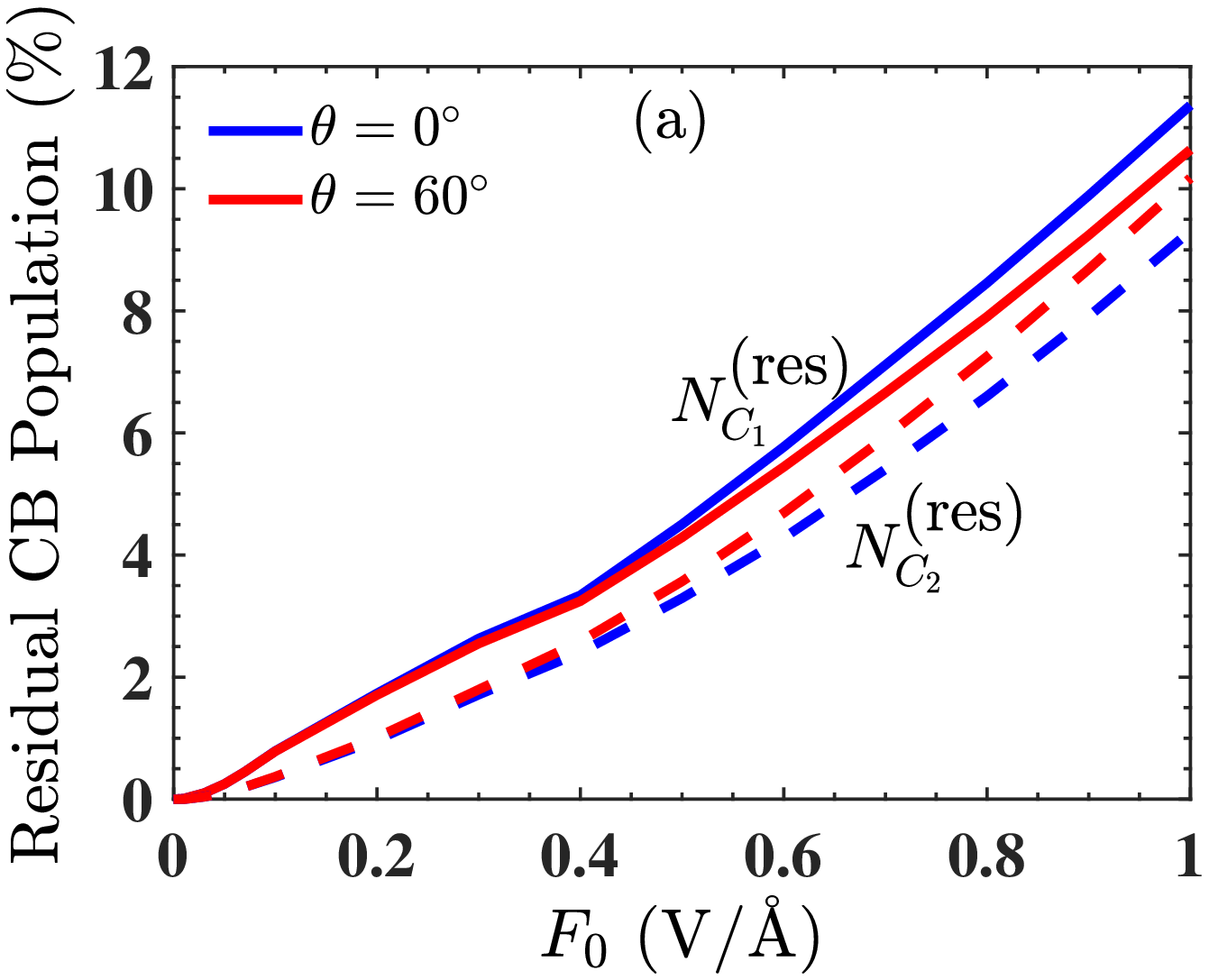}\\  
\includegraphics[scale=0.5]{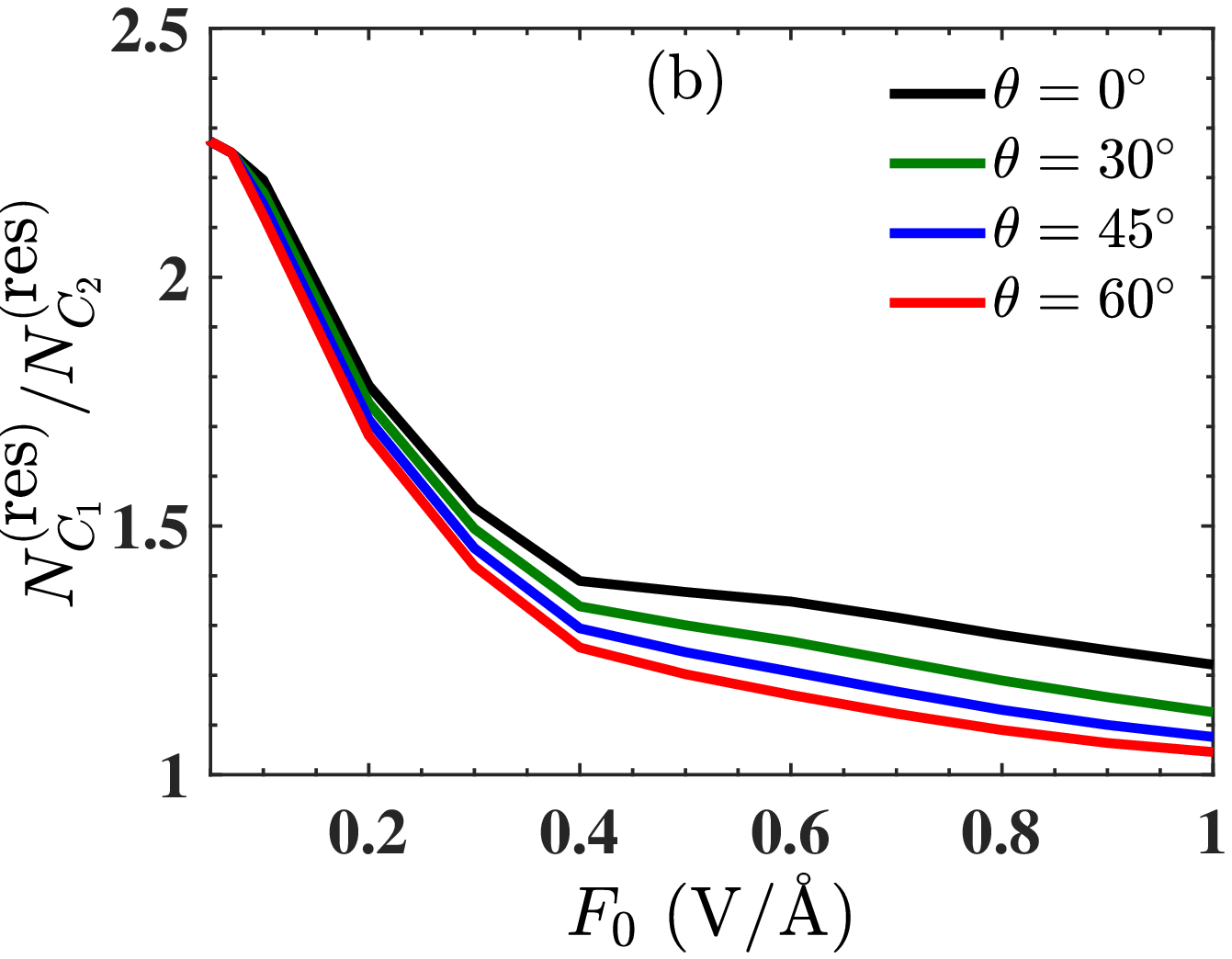} 
\end{tabular}
\end{center}
%\vspace*{-1cm}
\caption{(a) Residual population of the first, $N_{C_{1}}^{\mbox{(res)}}$ and the second, $N_{C_{2}}^{\mbox{(res)}}$, CB as a function of the pulse amplitude, $F_{0}$ and for different angles of incidence. Here solid and dashed lines correspond to $N_{C_{1}}^{\mbox{(res)}}$ and $N_{C_{2}}^{\mbox{(res)}}$, respectively. (b) Ratio of the residual CB populations, $N_{C_{1}}^{\mbox{(res)}}/N_{C_{2}}^{\mbox{(res)}}$ as a function of the field amplitude for different angles of incidence.}
\label{fig14}
\end{figure}

The residual CB populations of the first and the second CBs monotonically increase with the field amplitude, see Fig. \ref{fig14}(a). 
At the field amplitude of $F_{0}=1$ V/$\mbox{\AA}$, the CB population of the first CB reaches 12 $\%$, while the CB population of the second CB is about 9 $\%$. As expected, the CB population of the second CB is always less than the CB population of the first CB. This difference, which can be characterized as a ratio $N_{C_{1}}^{\mbox{(res)}}/N_{C_{2}}^{\mbox{(res)}}$ is shown in  Fig. \ref{fig14}(b). At small field amplitude this ratio is about 2.3. With increasing the field amplitude the CBs become more equally populated, which is due to the fact that electrons are excited from the VBs over the energy range that is proportional to the field amplitude.
\section{Conclusions}
\label{conclusion}
The electron dynamics in AB-stacked bilayer graphene is sensitive to the angle of incidence of the optical pulse. Such sensitivity is due to the fact that the normal component of the electric field of the pulse breaks the interlayer symmetry and, correspondingly, the inversion symmetry of bilayer graphene. As a result, for bilayer graphene in such optical field, only the $y$-axis is the axis of symmetry but not the $x$-axis.  
Braking of inversion symmetry in bilayer graphene opens a dynamic band gap. When the plane of incidence of the pulse is $x-z$ plane, such band gap results in nontrivial topological phase that behaves differently above and below the $K$ ($K^\prime$) point. Here the topological phase is defined as the sum of the geometric (Berry) phase and the phase of the interband dipole matrix element. Competition  between the dynamic phase and the nontrivial topological phase results in a topological resonance, which has different strength above and the below the $K$ ($K^\prime$) point. 
This behavior 
results in two effects: (i) the conduction band population distribution in the reciprocal space is asymmetric with respect to the plane of incidence of the pulse, $x-z$ plane, (ii) there is a nonzero transverse electric current generated during the pulse in the direction perpendicular to the plane of incidence, i.e., in $y$ direction. These features occur only for bilayer graphene with the bandgap, which, in our case, is the dynamic band gap generated by the normal component of the field. Thus, for the normally incident optical pulse, the conduction band population distribution is symmetric with respect to the plane of polarization of the pulse and the transverse electric current is exactly zero. Furthermore, if the plane of incidence of the pulse is the $y-z$ plane, which is still the plane of symmetry of bilayer grahene even in the optical field, no transverse current is generated in the system during thr pulse, even for oblique incidence.

\begin{acknowledgements}
Major funding was provided by Grant No. DE-FG02-11ER46789 from the Materials Sciences and Engineering Division of the Office of the Basic Energy Sciences, Office of Science, U.S. Department of Energy. Numerical simulations have been performed using support by Grant Nos. DE-FG02-01ER15213 and DE-SC0007043 from the Chemical Sciences, Biosciences and Geosciences Division, Office of Basic Energy Sciences, Office of Science, US Department of Energy.  Supplementary funding came from (i) Grant No. N000-14-17-1-2588 from the Office of Naval Research (ONR), (ii) the subaward no. 24086151 of the Grant No. FA9550-15-1-0037 from University of Central Florida, subcontracted by the Air Force Office of Scientific Research (AFOSR), and (iii) the subaward no.  T883032 of the Grant No.  EFMA-1741691 from Emory University, subcontracted by the National Science Foundation (NSF). PK has been supported in part by the Grant No. CMMI 1661618 from the National Science Foundation.
\end{acknowledgements}
%\begin{acknowledgements}
%Major funding was provided by Grant No. DE-FG02-11ER46789 from the Materials Sciences and Engineering Division of the Office of the Basic Energy Sciences, Office of Science, U.S. Department of Energy. Numerical simulations have been performed using support by Grant No. DE-FG02-01ER15213 from the Chemical Sciences, Biosciences and Geosciences Division, Office of Basic Energy Sciences, Office of Science, US Department of Energy.  Supplementary funding came from (i) Grant No. N000-14-17-1-2588 from the Office of Naval Research, (ii) 
%\end{acknowledgements}

\appendix
\section{Matrix elements of the non-Abelian Berry connection}
\label{appendixA}
The matrix elements of the in-plane component of the non-Abelian Berry connection given by Eq. (\ref{eq13}) has the following form
\begin{align}
\mathcal{A}^{v_{2}v_{1}}_{x}&=\mathcal{A}^{c_{2}c_{1}}_{x}=-\frac{2a\mathcal{N}_{v_{2}}\mathcal{N}_{v_{1}}\mathcal{U}_{1}}{|f(\mathbf{k})|^{2}}\;,\label{eqA1}\\
\mathcal{A}^{c_{1}v_{1}}_{x}&=-\mathcal{A}^{c_{2}v_{2}}_{x}=-\frac{2ia\gamma_{1}\mathcal{N}_{c_{1}}\mathcal{N}_{v_{1}}\mathcal{U}_{2}}{\mathcal{F}|f(\mathbf{k})|^{2}}\;,\label{eqA2}\\
\mathcal{A}^{c_{2}v_{1}}_{x}&=-\frac{2a\mathcal{N}_{c_{2}}\mathcal{N}_{v_{1}}\mathcal{U}_{1}}{|f(\mathbf{k})|^{2}}\;,\label{eqA3}\\
\mathcal{A}^{c_{1}v_{2}}_{x}&=-\frac{2a\mathcal{N}_{c_{1}}\mathcal{N}_{v_{2}}\mathcal{U}_{1}}{|f(\mathbf{k})|^{2}}\;,\label{eqA4}\\
\mathcal{A}^{v_{1}v_{1}}_{x}&=\mathcal{A}^{c_{2}c_{2}}_{x}=\mathcal{A}^{v_{2}v_{2}}_{x}=\mathcal{A}^{c_{1}c_{1}}_{x}=\frac{a\mathcal{U}_{1}}{|f(\mathbf{k})|^{2}}\;,\label{eqA5}\\
\mathcal{A}^{v_{2}v_{1}}_{y}&=\mathcal{A}^{c_{2}c_{1}}_{y}=\frac{2a\mathcal{N}_{v_{2}}\mathcal{N}_{v_{1}}\mathcal{U}_{3}}{\sqrt{3}|f(\mathbf{k})|^{2}}\;,\label{eqA6}\\
\mathcal{A}^{c_{1}v_{1}}_{y}&=-\mathcal{A}^{c_{2}v_{2}}_{y}=-\frac{2i\sqrt{3}a\gamma_{1}\mathcal{N}_{c_{1}}\mathcal{N}_{v_{1}}\mathcal{U}_{4}}{\mathcal{F}|f(\mathbf{k})|^{2}}\;,\label{eqA7}\\
\mathcal{A}^{c_{2}v_{1}}_{y}&=\frac{2a\mathcal{N}_{c_{2}}\mathcal{N}_{v_{1}}\mathcal{U}_{3}}{\sqrt{3}|f(\mathbf{k})|^{2}}\;,\label{eqA8}\\
\mathcal{A}^{c_{1}v_{2}}_{y}&=\frac{2a\mathcal{N}_{c_{1}}\mathcal{N}_{v_{2}}\mathcal{U}_{3}}{\sqrt{3}|f(\mathbf{k})|^{2}}\;,\label{eqA9}\\
\mathcal{A}^{v_{1}v_{1}}_{y}&=\mathcal{A}^{c_{2}c_{2}}_{y}=\mathcal{A}^{v_{2}v_{2}}_{y}=\mathcal{A}^{c_{1}c_{1}}_{y}=-\frac{a\mathcal{U}_{3}}{\sqrt{3}|f(\mathbf{k})|^{2}}\;.\label{eqA10}
\end{align}
where
\begin{align}
\mathcal{U}_{1}&=\sin\left(\frac{ak_{x}}{2}\right)\sin\left(\frac{\sqrt{3}ak_{y}}{2}\right)\;,\label{eqA11}\\
\mathcal{U}_{2}&=\sin(ak_{x})+\sin\left(\frac{ak_{x}}{2}\right)\cos\left(\frac{\sqrt{3}ak_{y}}{2}\right)\;,\label{eqA12}\\
\mathcal{U}_{3}&=\cos(ak_{x})-\cos\left(\frac{ak_{x}}{2}\right)\cos\left(\frac{\sqrt{3}ak_{y}}{2}\right)\;,\label{eqA13}\\
\mathcal{U}_{4}&=\cos\left(\frac{ak_{x}}{2}\right)\sin\left(\frac{\sqrt{3}ak_{y}}{2}\right)\;,\label{eqA14}\\
\mathcal{N}_{v_{1}}&=\mathcal{N}_{c_{2}}=\frac{\gamma_{0}|f(\mathbf{k})|}{\sqrt{2\left(E_{c_{2}}^{2}+\gamma_{0}^{2}|f(\mathbf{k})|^{2}\right)}}\;,\label{eqA15}\\
\mathcal{N}_{v_{2}}&=\mathcal{N}_{c_{1}}=\frac{\gamma_{0}|f(\mathbf{k})|}{\sqrt{2\left(E_{c_{1}}^{2}+\gamma_{0}^{2}|f(\mathbf{k})|^{2}\right)}}\;,\label{eqA16}
\end{align}

and $\mathcal{F}=\sqrt{4\gamma_{0}^{2}|f(\mathbf{k})|^{2}+\gamma_{1}^{2}}$.

Also, the matrix elements of the normal component of the non-Abelian Berry connection  can be written as
\begin{align}
\mathcal{A}_{z}^{c_{2}c_{1}}&=2L_{z}\mathcal{N}_{c_{2}}\mathcal{N}_{c_{1}}\;,\label{eqA17}\\
\mathcal{A}_{z}^{v_{2}v_{1}}&=2L_{z}\mathcal{N}_{v_{2}}\mathcal{N}_{v_{1}}\;,\label{eqA18}\\
\mathcal{A}_{z}^{c_{2}v_{1}}&=L_{z}\mathcal{N}_{c_{2}}\mathcal{N}_{v_{1}}\Big(1-\frac{E_{c_{2}}^{2}}{\gamma_{0}^{2}|f(\mathbf{k})|^{2}}\Big)\;,\label{eqA19}\\
\mathcal{A}_{z}^{c_{1}v_{2}}&=L_{z}\mathcal{N}_{c_{1}}\mathcal{N}_{v_{2}}\Big(1-\frac{E_{c_{1}}^{2}}{\gamma_{0}^{2}|f(\mathbf{k})|^{2}}\Big)\;,\label{eqA20}
\end{align}
and $\mathcal{A}_{z}^{c_{2}v_{2}}=\mathcal{A}_{z}^{c_{1}v_{1}}=0$.

\section{Matrix elements of the intraband velocity}
\label{appendixB}
From the known eigenstates, the matrix elements of the intraband velocities can be expressed in the following form:
\begin{align}
V_{x}^{c_{2}c_{2}}&=-V_{x}^{v_{1}v_{1}}=-\frac{2a\gamma_{0}^{2}\mathcal{U}_{2}}{\hbar\mathcal{F}}\;,\label{eqB1}\\
V_{y}^{c_{2}c_{2}}&=-V_{y}^{v_{1}v_{1}}=-\frac{2\sqrt{3}a\gamma_{0}^{2}\mathcal{U}_{4}}{\hbar\mathcal{F}}\;,\label{eqB2}
\end{align}
The other intraband velocities can be expressed as $V_{x}^{c_{1}c_{1}}=-V_{x}^{v_{2}v_{2}}=-V_{x}^{v_{1}v_{1}}$ and  $V_{y}^{c_{1}c_{1}}=-V_{y}^{v_{2}v_{2}}=-V_{y}^{v_{1}v_{1}}$.

\end{document}